# Ring gap resonant modes on disk/film coupling system caused by strong interaction

Guodong Zhu[#], Linhong Qv[#], Yangzhe Guo and Yurui Fang*

 *Key Laboratory of Materials Modification by Laser, Electron, and Ion Beams (Ministry of Education); School of Physics, Dalian University of Technology, Dalian 116024, P.R. China.*

\# These authors contributed equally.

\* Corresponding emails: yrfang@dlut.edu.cn (Y. Fang)

**Abstract:** Peculiar ring gap modes on the surface of disk close to the metallic thin film are excited in the visible light regime. We apply plasmon hybridization method to illustrate the ring gap modes arising from the interaction between localized disk plasmons and continuum surface plasmons, which cannot be easily excited by the plane wave with polarization parallel to the film interface. In the coupled system, the hybrid modes energy and the surface charge distribution of nanoparticle are investigated both in simulation and hybridization method, showing consistence with each other. The excitation of ring gap modes provides further insight into strong coupling of the plasmon and the design of novel nanostructures.
**PACS:**

   Surface plasmon gap modes provide great potential applications on surface enhancement and strong interactions in varies researches and systems because metal nanostructures have the ability to localize the electromagnetic energy on the interface extremely. The interactions between the discrete, localized plasmons and the continuum, delocalized plasmons are really complex and may lead to some anomalous electromagnetic modes. The unique properties of interaction between localized surface plasmons (LSPs) and surface plasmon polaritons (SPPs) on particle-film systems have attracted a lot of attention in the nanoscience [1-5], enhanced interactions[6-10] and others[11]. Small symmetric structures like nano disks or spheres usually cannot support ring modes or breathing modes under plane wave excitation with normal illumination because of the symmetry, though the ring modes on such symmetric structures have been investigated with electron beam excitation [12-14]. In a coupled particle/film system, the interactions can be expanded with symmetric/antisymmetric ring modes on particles and films for energy level hybridization[15,16]. However, such modes never be observed in simulations or experiments with plane wave illumination.

   When ring electromagnetic modes including higher order modes are excited, which show more wave nodes on the surface of the disk (like cylindrical surface waves on the water), in sharp contrast with the dipolar and quadrupolar mode parallel to the polarization direction of the incident light excited on a single nanodisk. Such ring modes on small particles below 100 nm have very large wave vector which are almost impossible excited in visible region. However, in particle/film system, the strong coupling will bring in the hybridization of the large wave vector of particle with the low wave vector modes on film, which makes it possible observe the ring modes in visible region. In such system, the ring modes which may be call ring gap resonance modes, will bring profound understanding and applications in strong interactions.

In this letter, we apply plasmon hybridization (PH) method to calculating the plasmon energies of a disk near a metallic film and demonstrating the existence of ring gap resonance modes on disk/film coupling system (Fig. 1). The dependence of plasmon resonance energies and surface charge distributions on geometric parameter of disk and film are verified with the finite element method (FEM) to further explore the underlying properties of the ring gap resonance modes. In Fig 1(b), we can see the surprising ring gap mode of the disk/film system. Differing from the traditional symmetric ring surface charge distribution excited by point dipole or electron beam, the surface charge excited by a normal incident plane wave exhibits a special antisymmetric distribution owning to the present of metallic film. The study have potential applications in strong coupling and fundamental researches.

A simple and intuitive description of the plasmon resonance in composite nano-objects was presented by P. Nordlander et al [17,18], which expresses the plasmon resonances of a complex nanostructures as the interaction between plasmon resonances of simpler nanoparticles[19]. Here, the hybridization of metallic disk and surface is derived. For simplicity, the plasmon modes on the disk/film system can be expressed as the interaction of the surface plasmons of metallic surface and plasmons on the lower disk surface. In PH method, the plasmons are treated as incompressible irrotational deformations of the conduction electron gas in the particle[17]. We assume a uniform conduction electron density $n_d$ in the disk and $n_f$ in the film corresponding to bulk plasmon energies $\omega_{dB} = \sqrt{4\pi n_d e^2/m_e}$ and $\omega_{fB} = \sqrt{4\pi n_f e^2/m_e}$. The interaction potentials between the surface plasmons can be calculated by the instantaneous Coulomb potential.

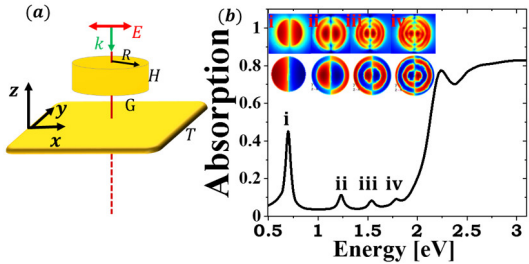

FIG 1. (a) The schematic depicting the geometrical parameters of disk/film system. The geometric variables $R$, G, $T$, $H$ represent the disk radius, the disk-surface separation, the metallic film thickness, and the disk thickness. (b) The optical absorption spectrum of disk/film system simulated with FEM when $R = 50\ nm$, $H = 50\ nm$, and the gap $G = 1\ nm$. Here, the film is thick enough to be as a metallic surface. The absorption peak positions have been marked as ( i - iv ). The symmetric surface electric field distribution and the antisymmetric surface charge distribution on nano disk corresponding to the absorption peaks are inserted in (b).

The potential on the lower surface of disk can be expressed as[20]

$$\Phi(r,z,\emptyset,t) = \sum_{k_1,m} \Phi_{k_1 m} = \sum_{k_1,m} A_{k_1 m} J_m(k_1 r) \cos(m\emptyset) e^{k_1(z-z_0)} \quad (1)$$

where the index $k_1$ is non-negative as is the azimuthal index $m$, $A_{k_1 m}$ is the non-retarded mode time-dependent amplitude, and $z_0$ is the distance between the lower surface of disk and metallic surface, and the metallic surface is located at the plane $z = 0$ of Cylindrical coordinates. One finds the charge density $\rho$ corresponding to the potential of equation (1) satisfies Poisson's equation, so the potential energy can be given by the integral $U = \frac{1}{2}\int \rho \Phi dV = \frac{1}{2}\sum_{k_1,m}|A_{k_1 m}|^2 N_{k_1 m}^2 k_1$, $N_{k_1 m}^2 = \int_0^R J_m(k_1 r) J_m(k_1' r) r dr$, the upper limit of integration is the radius of disk. The kinetic energy $T$ of the oscillations is given by the volume integral $T = \frac{m n_d}{2}\int \dot{\mathbf{D}}\cdot\dot{\mathbf{D}} dV$ and the charge displacement vector is $\mathbf{D} = -\frac{e}{m}\sum_{k_1,m}\frac{1}{\omega_{k_1 m}^2}\nabla\Phi_{k_1 m}$. One finds $T = \frac{1}{2}\sum_{k_1,m}[\frac{\dot{A}_{k_1 m} N_{k_1 m}}{\omega_{k_1 m}}]^2 k_1$. The total Lagrangian describing the plasmon dynamics on the disk is

$$L_{disk} = \frac{1}{2}\sum_{k_1,m}\left([\dot{A}_{k_1 m}]^2 - \omega_{k_1 m}^2 [A_{k_1 m}]^2\right) N_{k_1 m}^2 k_1/\omega_{k_1 m}^2 \quad (2)$$

The eigenmode frequency $\omega_{km}$ satisfies the dispersion relation $\omega_{k_1 m}^2 = \frac{\omega_{dB}^2 I_m'(k_1 z_0) K_m(k_1 z_0)}{I_m'(k_1 z_0) K_m(k_1 z_0) - I_m(k_1 z) K_m'(k_1 z_0)}$. The potential of metallic surface is $\Phi_{surface} = \sum_{k_2,m} B_{k_2 m} J_m(k_2 r) \cos(m\emptyset) e^{-k_2 z}$. Then the Lagrangian of metallic surface can be expressed as

$$L_{surface} = \frac{1}{\omega_{fB}^2}\sum_{k_2,m}\left([\dot{B}_{k_2 m}]^2 - \omega_{k_2 m}^2[B_{k_2 m}]^2\right) \quad (3)$$

The energy of the plasmon resonance corresponding to a flat is $\omega_{k_2 m} = \omega_{fB}/\sqrt{2}$. The interactions between the disk and metallic surface plasmons are given by

$$V_I = \frac{1}{2}\sum_{k,m}\int \rho_{k_1 m}\Phi_{surface} dS \quad (4)$$

where the potential energy is given by the integral on lower surface of the disk. The integral can be evaluated analytically and takes the form

$$V_I = \frac{1}{2}\sum_{k_1}\sum_{k_2} k_1 e^{-k_2 z_0} A_{k_1 m} B_{k_2 m} \int_0^R J_m(k_1 r) J_m(k_2' r) r dr \quad (5)$$

After rescaling the degrees of freedom $\sqrt{\frac{k_1}{\omega_{dB}^2}N_{k_1m}^2}A_{k_1m} \to A_{k_1m}$ and $\sqrt{\frac{1}{\omega_{fB}^2}}B_{k_2m} \to B_{k_2m}$, the total Lagrangian is written as $L = \sum_m L^m$, where $L^m$ takes the following form

$$L^m = \sum_{k_1}\left([\dot{A}_{k_1m}]^2 - \omega_{k_1m}^2[A_{k_1m}]^2\right) + \sum_{k_2}[\dot{B}_{k_2m}]^2 - \omega_{k_2m}^2[B_{k_2m}]^2 \tag{6}$$
$$- \frac{1}{2}\omega_{dB}\omega_{fB}\sum_{k_1}\sum_{k_2}\frac{\sqrt{k_1}e^{-k_2z_0}}{N_{k_1m}}\int_0^R J_m(k_1r)J_m(k_2'r)rdr\, A_{k_1m}B_{k_2m}$$

Through solving the Euler-Lagrange equations, the plasmon energies can be obtained. In the calculations presented above, we use 1000 $k$-points from $10^{-3}$ to 1 $nm^{-1}$ and the wave vector on the lower surface of disk satisfy $J_m'(k_1R) = 0$ devivated from the limiting condition for the modes of a cylinder $\frac{\epsilon_{metal}}{\epsilon_{dielectric}} = \frac{I_m(k\rho_0)K_m'(k\rho_0)}{K_m(k\rho_0)I_m'(k\rho_0)} \approx -1$[21].

For one $m$ value, there will be different eigenvalue corresponding the n$^{th}$ zero nodes, which can be expressed as $(m, n)$ for the eigenmodes. With the matrix calculation developed from plasmon hybridization[17], one can achieve the characteristics of different complex nanostructure quickly.

The energy shift of the plasmon resonances for a lower surface of disk near an interface of semi-infinite bulk material is shown in Fig 2. Because of the symmetry under linear polarized plane wave excitation, the even modes (including $m = 0$) cannot be excited. Panel (a) shows Plasmon energy of the modes when $m = 1$ while panel (b) refers to $m = 3$. The upper curves describe the plasmon energy of anti-bonding mode and the lower curves are related to bonding mode. Plasmon energy of anti-bonding and bonding modes show a smaller gap with the increasing gap distance, which is attributed to the weaker coupling between disk and metallic surface. The bonding mode corresponding to $(m, 1)$ modes show a large separation with other modes and the other modes tend to overlap when the distance $z_0$ is far enough, which is consistent with the FEM result in Fig 1b. In addition, though $(m, 2)$ modes have the tendency to overlap with other higher modes, it may be visible at about 2 eV ($m = 1$). In Fig. 2b, the $(3, n)$ modes have a higher plasmon energy than $(1, n)$ modes. In Fig. 2c, the imaginary part of the dipole polarizability of the plasmon resonances is plotted when $m = 1$ and $z_0 = 3\, nm$. The optical absorption is thus characterized by several peaks located around 2.07 eV, the weak features around 2.4 eV are the antibonding plasmons resulting from the interaction of the surface plasmons and the disk plasmons. Moreover, the relative position of energy peaks in Fig. 2(c) is consistent with the relative position of the absorption peaks in Fig. 1(b).

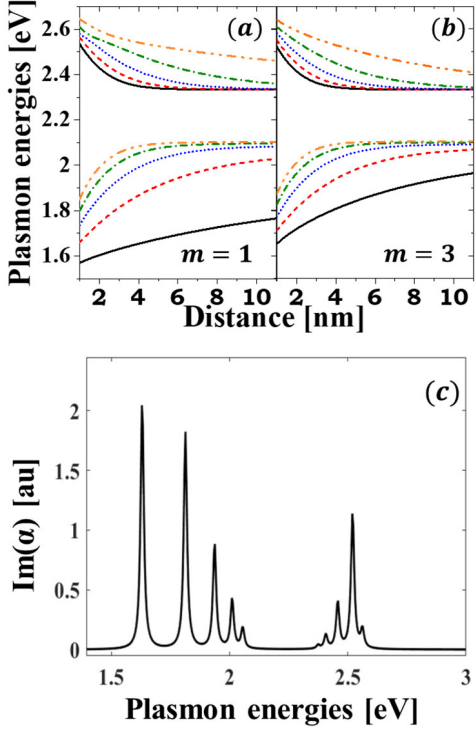

FIG 2. The energy shift of the five disk plasmon modes as a function of distance $G$ between the disk and metallic surface with bulk plasmon resonance energy $\omega_{fB} = 3.3$ eV. The radius of disk $R$ is $50\ nm$. Panel (a) refers to $m = 1$ and panel (b) to $m = 3$. The lines are related to the wave vector of first five zero nodes of $J'_m(k_1 R)$, corresponding to solid (black), dash (red), short dot (green), short dash dot (blue) and double-dotted (yellow) respectively, which are the result of PH. Each color corresponds to two curves related to the anti-banding and banding mode of plasmon hybridization. (c) The imaginary of the dipole polarizability $\alpha(\omega + i\delta)$ (in arbitrary units) of $m = 1$ as a function of plasmon resonance frequency. We assume $\delta = -0.008$ to broaden the width of absorption spectrum. The arbitrary units for polarizability are used in here to show the relative contributions of the plasmon modes related to different $k_1$.

To better understand the physical mechanism of the ring gap resonance modes, the finite element method is used to simulate the size dependent resonance of the coupling system in detail. Firstly, we explore the effects of changing the disk radius when the disk and film are thick (50 nm). In this case, the plasmonic density of states is a delta function around the $\omega_{k_2 m} = \omega_{fB}/\sqrt{2}$. We can ignore the interactions between the upper and lower surfaces of disk and film, respectively, which is consistent with the above theoretical assumption. Therefore, as shown in Fig. 3, the availability of the above theory elaborating ring gap modes can be verified through the simulation. Also, some more high-order resonance modes were discovered on larger disks. When the diameter of the disk is larger and the film is thinner, the symmetric electric field distribution and anti-symmetric surface charge distribution is depicted in Fig. 4. The charge distributions imply that the resonances of the disk are anti-symmetric ring modes, which cannot be directly excited by the planewave, but here is caused by the hybridization of the disk and surface.

To check if the thickness of film always has an effect on the hybridization of disk/film system, the thickness dependence of film is investigated as shown in Fig. 5. With the increasing thickness of film, the effective continuum of film trend to lie at higher energies, one can find the resonances will blue shift resulting from the weaker hybridization between the upper and lower surface of film. Finally, the interactions of disk and film that are as thin as 10 nm are also investigated to make a thorough inquiry of ring gap mode in disk/film system.

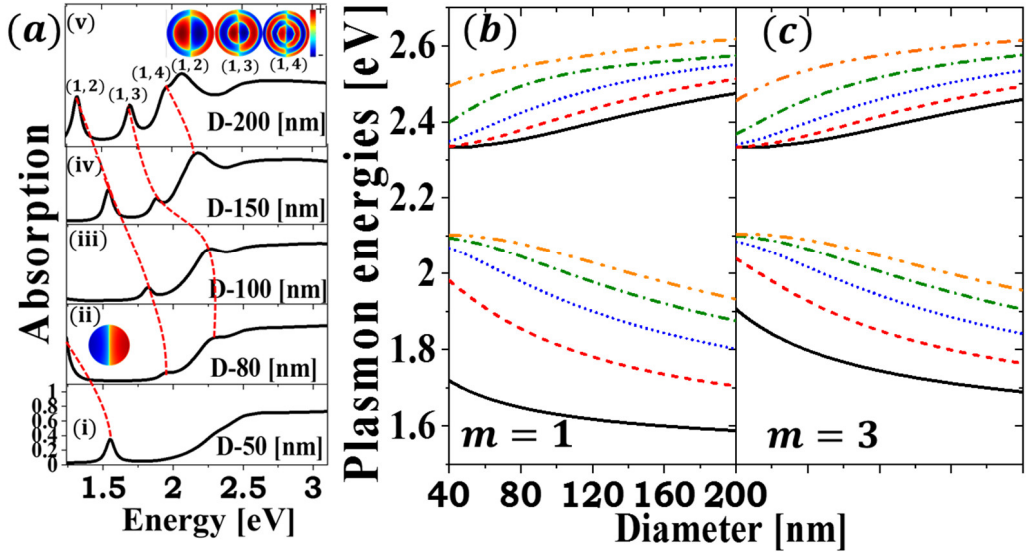

FIG 3. (a) The absorption spectra of the disk with a varying diameter from 50 nm to 200 nm and a height of 50 nm. The gap between Au disk and Au film is 3 nm. The film is thick enough to be seen as a metallic surface. The red dotted line in the figure marks the change in absorption peak positions. The inset of (a-V) shows the electric field distributions of hybrid modes that have been marked. (b) and (c) shows the plasmon energies of disk/film system with different disk diameters calculated by the PH matrix method.

Figure 3 shows the numerical simulation result of thick disk-film systems. When the diameter of disk is small, there is one obvious absorption peak at low energy, arising from the dipole excited on the lower surface of disk (inset in Fig. 3a-ii). And two modes $(1, 2)$ and $(1, 3)$ arises as marked in Fig. 3a (also see insets in Fig. 3a-V). With the increasing of diameter, the excited modes red shift. The peaks are not obvious when the diameter of disk is 80 nm, and become distinctive as the diameter increases. Meanwhile a new resonance mode $(1, 4)$ emerges. We can attribute this to the hybridization of disk and film, the new excited high-order modes with large SPP wave vector (almost 10 times as the wave vector on plane) take part in the hybridization, giving rise to the emergence of plasmon ring resonance peaks. The peak close to peak 3 for D = 200 nm is arising from the coupling of the edge mode of the disk. As disk size increasing, more high-order modes will appear, whose interaction with the plasmon on plane surface will lead to the lower modes red shift further. Figure 3(b) shows the result of hybridization calculation. We can see when the radius is small, the high-order energies show little difference and the energy of (1, 1) is especially obvious. As the radius increases, the energies will red shift and high-order modes trend to be visible without mixing with each other. For $m = 3$ modes, the hybridization is very similar to $m = 1$, but with higher energy as shown in Fig. 3c.

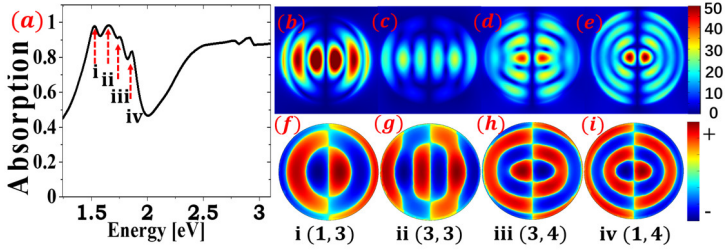

FIG 4. (a) the optical absorption spectrum of the disk with $D = 200\ nm$, $H = 50\ nm$, $T = 10\ nm$ and $G = 3\ nm$. The corresponding surface charge distribution of absorption peaks are shown in the inset. (b-e) The electric field distributions correspond to the four absorption peaks.

In Fig. 4, A disk with a D = 200 nm diameter is placed on a thin film, the $m = 1$ and $m = 3$ modes are simultaneously excited (Fig. 4b-i). From the surface change distributions and electric field distributions shown in (b) to (i), we can assign peak i and iv to $m = 1$ modes and peak ii and iii to $m = 3$ modes, respectively. For $m = 1$ modes, peak i is hybrid (1, 3) and peak iv is hybrid (1, 4) ring gap mode, respectively. For $m = 3$ modes, peak ii is hybrid (3, 3) and peak iii is hybrid (3, 4) order ring gap modes. From the electric field distributions one can see that the field intensity distributions show feature of ring shape. Because the excitation light is plane wave and the polarization is linear along x, the electric field distributions show symmetric feature along the middle line in y direction. And consequently, the surface charge distributions in Fig. 4 (f-i) show anti-symmetry in left and right parts on disks, which shows that the ring gap modes are not breathing modes. The energy of the two modes for $m = 3$ is between the two modes of $m = 1$, which is not exactly matching the hybridization in Fig. 3c. In Fig. 4, thin film is used to replace the semi-infinity plane because it is easier to get $m = 3$ modes. For thin film, the bonding mode have lower energy than semi-infinity plane, so when the disk is coupling with the film, the energy of (3, 4) mode of disk is closer to the resonance of the film so as that their coupling is much stronger than (3, 3) mode of disk, which makes the hybrid (3, 4) mode have lower energy. Another interesting phenomenon is that in Fig. 4 (b-e), one can find that the electric fields in the center ring are stronger than the outer ring, which shows a concentration of the electromagnetic energy. Another one of important reason for the appearance of higher m modes is the coupling between particles from the periodic boundary condition.

For a deeper understanding on the origin of the ring gap modes, we now consider the case where the film thickness varies from 10 nm to 30 nm. It can be referred from Fig. 5 (a) that as the film thickness increases, the intensity of absorption peaks decrease significantly and the number of peaks gradually decrease resulting from the weaken coupling between the plasmon on lower surface of film and the modes on the other surfaces. When the film becomes thicker, the bonding energy levels on the upper and lower surfaces of the film will have higher energy that pushes the hybridization energy levels between the disk and the film to higher energy[22]. Meantime, the weakening of coupling also causes the blue shift of resonances energies in Figure 5(a), where the location of peaks has been labelled. As the film becomes thicker, the resonance peak intensity gradually decreases, which mainly reflects the coupling strength between the disk mode and the film, that is the lower surface of the film is less involved in hybridization[15].

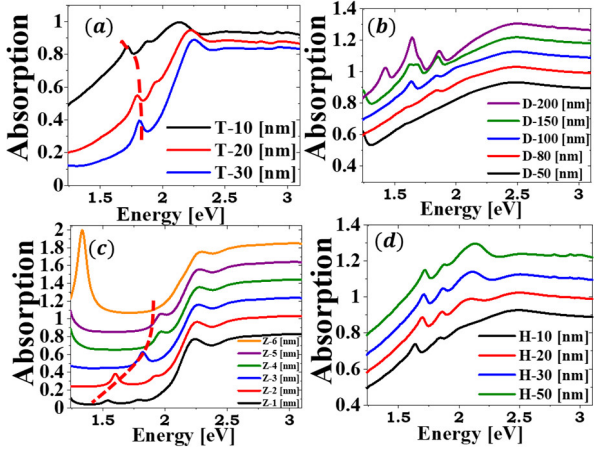

FIG. 5. (a) The optical absorption spectrum of the disk with a 50 nm height and a diameter with 100 nm. The gap between Au disk and Au film varying from 10 nm to 30 nm is 3 nm. The different curves correspond to the increasing thickness of film. (b) The absorption spectrum of disk and film with 3 nm gap, whose thickness are all 10 nm. The different curves correspond to the increasing diameters of disk. (c) The optical spectrum of the disk with 100 nm diameter and a height of 50 nm. The gap between Au disk and Au film varies from 1 nm to 7 nm. (d) The optical absorption spectrum of the disk with a 100 nm diameter and a varying thickness. The Au film thickness is 10 nm and the gap is 3 nm. In order to clearly distinguish the shape of spectral lines, each line of (b) and (d) has an offset of 0.1 and (c) has a 0.2 offset.

Figure 5 (b) exhibits the effect of changing the diameter of disk when the disk and the film thickness are all 10 nm. As analyzed above, the larger diameter means the more modes with lower energy on the disk can take part in the hybridization. So, the number of absorption peaks increases as the diameter increases. The $k_1$ is inversely proportional to the diameter and proportional to the energy of modes. In this case, the hybridization between the upper and lower surface of film is unignored; the effective surface plasmon continue has lower energy [15]. Compare with Figure 3 (a), the absorption peak intensity of the thinner disk is significantly increased. The resonance energy of modes on the disk decreases with the diameter increasing leading to weaker coupling of disk and film. As a result, the energy after hybridization will blue shift. Finally, the effect of the gap between disk and film and the thickness of disk is investigated, so that a comprehensive exploration of the disk/film system is carried out.

In Fig. 5c, the resonance energy shows a strong blue shift with the gap increasing, which is consistence with the calculated results in Fig. 2 via plasmon hybridization method. The distance dependent energy shift always contains image-like contribution, which will result in a blue shift of the energy with increasing distance. The increasing thickness of disk will cause the disk resonance blue shifting and consequently results in the blue shift of the system (Fig. 5d).

In conclusion, when the metal disk is placed on a metallic film, the hybridization between high order modes of disk which has large wave vector and the continuous surface plasmon state of the film will cause the energy shift of the high order ring modes, which cannot be excited in isolated disk with plane wave illuminate normally. With the theoretical PH analysis and the finite element method simulation, we show that the existence of ring gap modes and analyze the optical characteristics of disk/film system. If the ring modes are treated as standing waves, the wave vector will be much higher than the surface plasmon wave vector on a plane. In electron energy loss spectroscopy experiments, the symmetric ring modes on disk will have much higher resonance energy. And the ring modes on disks usually can only be excited with local point sources. Here when the disk is brought close to the plane with small gap, the strong coupling will greatly lower down the resonance energy of the ring modes but with anti-symmetric surface charge distributions in a ring. Such large surface plasmon wave vector in visible region is not common in metal SPPs. Similar phenomenon should be observed in other nanoparticle/film system like cube on film. We also show that the nature of the interaction is essential related to the film thickness, the disk thickness and the diameter of disk, through which we can tune the resonance of the ring gap modes. When the system is excited with circularly polarized light, the finally pattern will be continuous rings with nodes composed of two orthogonal polarization light with phase delay. This work gives a clear description of the plasmon resonance ring gap modes of a disk/film system. The results are really useful to the development of hybridization method and fundamental research in physics.

## Method
**Electrodynamics simulation**
Electromagnetic numerical simulations were performed with commercial finite element method (FEM, COMSOL Multiphysics 5.2). The model structure is a Au disk with diameter D (5 nm rounded edge) putting on a SiO2 dielectric layer (thickness is G) on a thickness T Au film (when T is thick enough, it performs as a bulk material). The surrounding medium was water. Periodic boundary conditions (300 nm) were used in horizontal direction and port boundary conditions were used for excitation in vertical direction. The excitation light was linearly polarized in x direction and incident from the particle side. The absorption was got from the reflection and transmission from ports.

## Funding
National Natural Science Foundation of China (NSFC) (12074054, 11704058); Fundamental Research Funds for the Central Universities (DUT19RC(3)007).

**Acknowledgement**

**Disclosures**
The authors declare no conflicts of interest.